\documentclass[fleqn,twoside]{article}
\usepackage[headings]{espcrc2}
\usepackage{epsfig,amsfonts}
\usepackage{graphicx}
\usepackage{dcolumn}
\usepackage{bm}
\DeclareMathSymbol{\lesssim}      {\mathrel}{AMSa}{"2E}
\DeclareMathSymbol{\gtrsim}       {\mathrel}{AMSa}{"26}
\def\be{\begin{equation}}
\def\ee{\end{equation}}
\def\bc{\begin{center}}
\def\ec{\end{center}}
\def\bea{\begin{eqnarray}}
\def\eea{\end{eqnarray}}

\def\nn{\nonumber}
\title{Two-loop QCD corrections to the vector form factors for the 
heavy-quark photo-production}

\author{R.~Bonciani\address[Freiburg]{Physikalisches Institut, 
        Albert-Ludwigs-Universit\"at Freiburg,
        D-79104 Freiburg, Germany}%
        \thanks{This work was supported by the European Union under
        contract HPRN-CT-2000-00149.}}

\begin{document}

\begin{abstract}
We review on the calculation of the heavy-quark photo-production
vector form factors, with the full dependence on the mass of the 
heavy-quark. The Feynman diagrams are evaluated within the
dimensional regularization scheme and expressed in Laurent series of 
$\epsilon=(4-D)/2$, where $D$ is the space-time dimension.
The coefficients of the expansion are expressed in terms of harmonic 
polylogarithms. The numerical evaluation of the analytical formulas 
and the threshold limit of our results are presented.
\vspace{1pc}
\end{abstract}

\maketitle

The measurement of the forward-backward asymmetry of the production of 
heavy quarks in $e^+ e^- \to Q \bar{Q}$ processes, such as $A_{FB}^{c,b}$
on the $Z^0$ peak \cite{EWWorkingGroup}, and, at the next generation of 
Linear Colliders, also the measurement of $A_{FB}^{t}$ \cite{TOPexp},
is a stringent test of the Standard Model, providing a precise determination 
of $\sin^2 \theta_W^{eff}$ \cite{Yellow}. 

The present theoretical description includes the NNLO QCD corrections.
For massless quarks, they were calculated numerically in \cite{Altarelli} 
and analytically in \cite{vanNeerven}, while in \cite{Catani} the order
${\mathcal O}(\alpha_S^2)$-corrections were calculated numerically 
for the b-quark retaining
terms that do not vanish in the small-mass limit (constants and 
log-enhanced terms), but neglecting both terms containing linear 
mass corrections, like $m_b^2/s$, and terms in which such a ratio is enhanced 
by a power of $\log(s/m_b^2)$. In order to take into account 
also this kind of terms, a full analytic calculation in which the mass of the 
heavy quark is kept systematically different from zero is required.
Three classes of contributions are involved: the tree level matrix elements 
for the decay of a vector boson into four partons, at least two of which 
being the heavy quark-antiquark pair; the one-loop corrected matrix elements 
for the decay of a vector boson into a heavy quark-antiquark pair plus a gluon;
the two-loop corrections to the decay of a vector boson into a heavy 
quark-antiquark pair.

In this paper we consider the virtual NNLO QCD corrections to the process
$\gamma^{*} \to Q(p_1) \bar{Q}(p_2)$, with
$P=p_{1}+p_{2}$ the momentum of the virtual photon and $p_{1}$, $p_{2}$ 
the momenta of the outgoing on-shell quark and antiquark respectively
($p_{1}^{2}=p_{2}^{2}=-m^2$ with $m$ the mass of the heavy quark). 
The vertex amplitude $V^{\mu}(p_1,p_2)$ can be expressed in terms of 
two form factors as:
\bea
V^{\mu}(p_1,p_2) \! \!\! \! &=& \! \! \! \! \bar{u}(p_1) 
\Gamma^{\mu}(p_1,p_2) v(p_2), \\ 
\Gamma^{\mu}(p_1,p_2) \! \!\! \! &=& \! \!\! \! F_{1}(p^2)
\gamma^{\mu} 
\!  + \!\frac{1}{2m} F_{2}(p^2) \sigma^{\mu \nu} P_{\nu} \! , 
\eea
where $\bar{u}(p_1), v(p_2) $ are the spinor wave functions of the 
quark and antiquark, 
$\sigma^{\mu \nu} = - \frac{i}{2}[ \gamma^{\mu},\gamma^{\nu}]$, 
\begin{figure}
\centering
\vspace*{25mm}
\epsfig{file=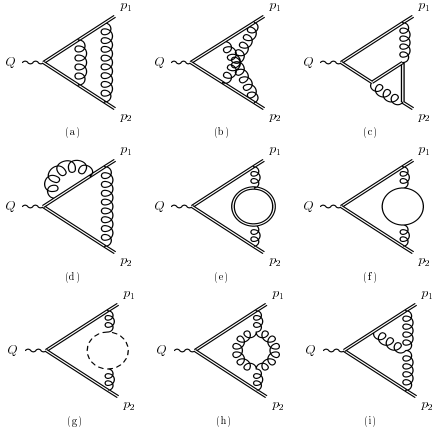,height=3.6cm,width=4.8cm,
        bbllx=160pt,bblly=250pt,bburx=460pt,bbury=500pt}
\vspace*{-6mm}
\caption{Two-loop Feynman diagrams. The curly lines represent gluons; the double 
straight lines, quarks of mass $m$; the single straight lines, massless quarks 
and the dashed lines ghosts.}
\label{fig1}
\end{figure}
and $p^2 = P^2/m^2 = -s/m^2$, with $s$ the squared c.m. energy.

At the two-loop level, the contributions to $F_{1}(p^2)$ and $F_{2}(p^2)$ 
come from the diagrams shown in Fig.~\ref{fig1}. They are evaluated in
\cite{us} using a method based essentially on two steps: 
\begin{enumerate}
\item By means of suitable projector operators, the contributions to 
$F_{1}(p^2)$ and $F_{2}(p^2)$ are expressed in terms of a big number of 
scalar integrals, whose ultraviolet (UV) and infrared (IR) singularities 
are regularized within the dimensional regularization scheme. Applying 
the so-called Laporta algorithm, introduced in \cite{Lap} and become, 
in the last few years, a standard technique for the calculation of 
Feynman diagrams, all these scalar integrals are expressed
in terms of 34 independent integrals called Master Integrals (MIs). 
The algorithm is based on Integration by Parts Identities \cite{Chet}, 
Lorenz invariance identities \cite{LI} and General Symmetry Relations.
It is completely algebraic and it is implemented in a code written in
FORM \cite{FORM}.
\item The calculation of the Master Integrals, then, is carried out 
\cite{RoPieRem1,RoPieRem3} by means of the Differential Equations 
technique \cite{DiffEq}.
\end{enumerate}
The renormalization of UV divergences is performed in a 
hybrid scheme: we renormalize the mass and wave function of the heavy
quark in the on-shell renormalization scheme, while the coupling
constant, as well as the light-fermion and gluon wave functions, are
renormalized in the $\overline{\rm MS}$ scheme.

The final result of the calculation is an analytical expression for the 
form factors, written in terms of harmonic polylogarithms of one variable
\cite{Polylog,Polylog3}. The result has still IR divergences that appear 
as poles in $1/\epsilon$. These divergences have to be canceled against 
those arising from the real radiation, which are not taken into 
account here.

\begin{figure}
\begin{center}
\includegraphics[height=52mm,angle=0]{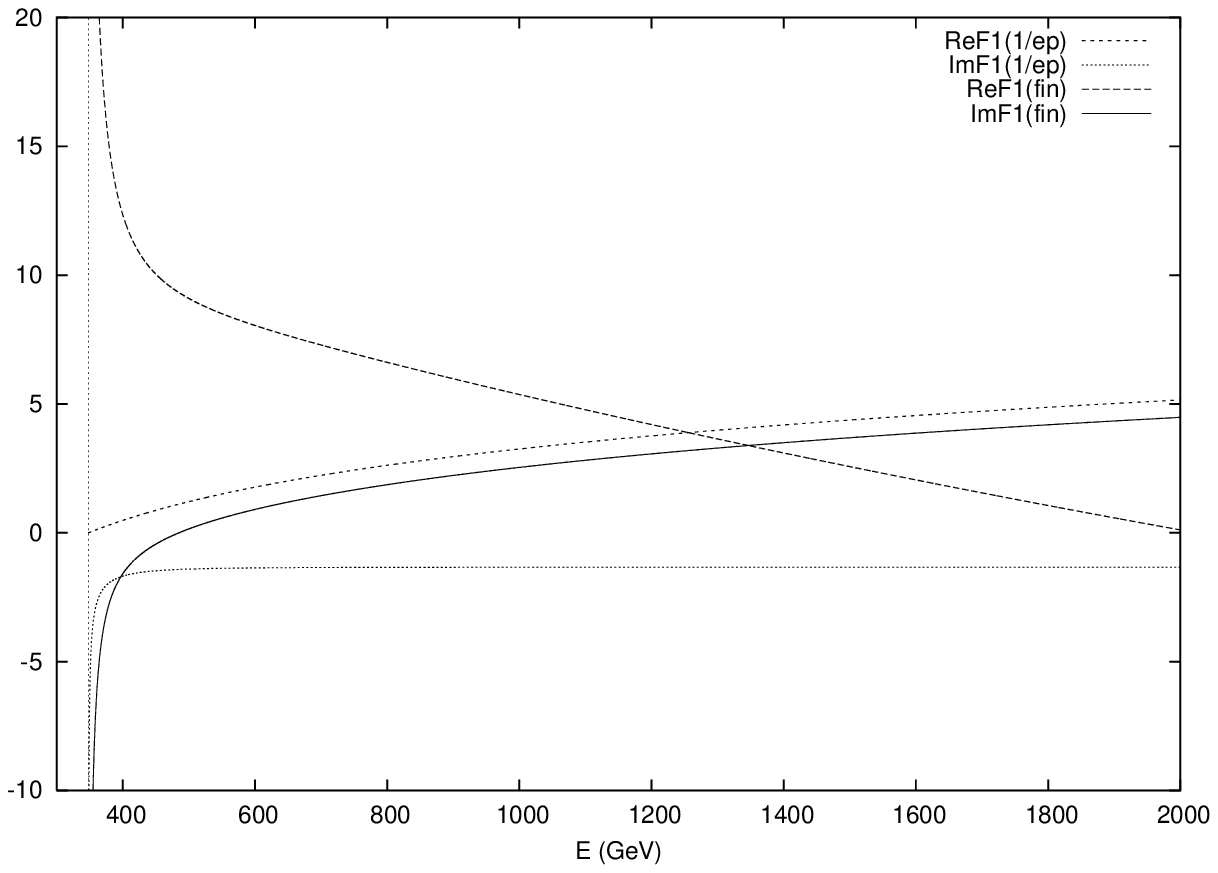}
\includegraphics[height=52mm,angle=0]{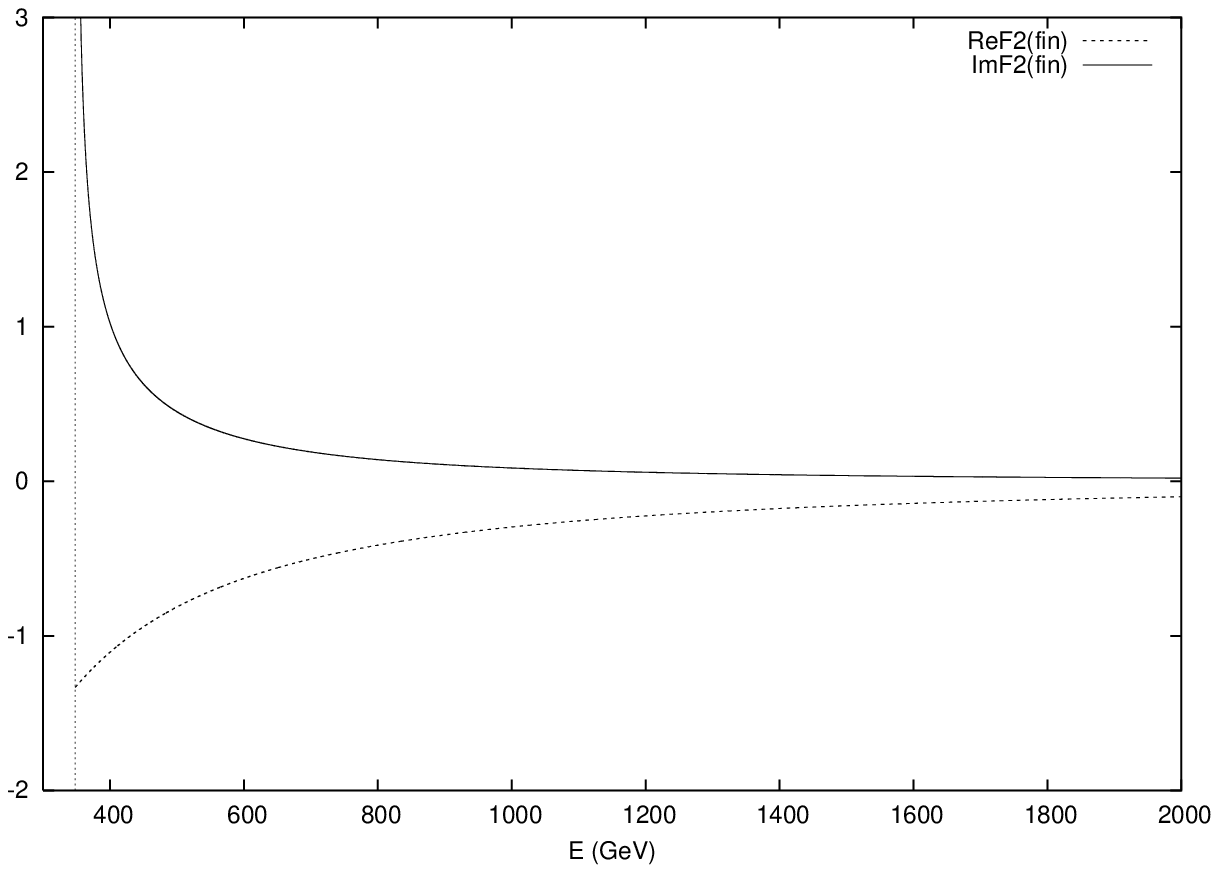}
\vspace*{-12mm}
\caption{One-loop form factors, $F_1$ and $F_2$, in the case of top-antitop 
production.The charge form factor (upper plot) has an IR pole, while the 
magnetic form factor is IR finite.
Note the Coulomb divergences of the form factors in the threshold limit.
The real part of the pole of $F_1$ and the real part of $F_2$ are, instead, 
finite (0 and -$C_{F}$ respectively).}
\label{fig2}
\vspace*{-8mm}
\end{center}
\end{figure}

\begin{figure}
\begin{center}
\includegraphics[height=52mm,angle=0]{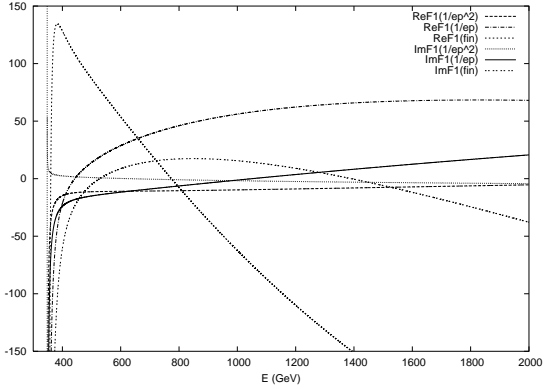}
\vspace*{-12mm}
\caption{Two-loop charge form factor, $F_1$, in the case of top-antitop 
production. The charge form factor has a double IR pole. }
\label{fig3}
\end{center}
\end{figure}

\begin{figure}
\begin{center}
\includegraphics[height=52mm,angle=0]{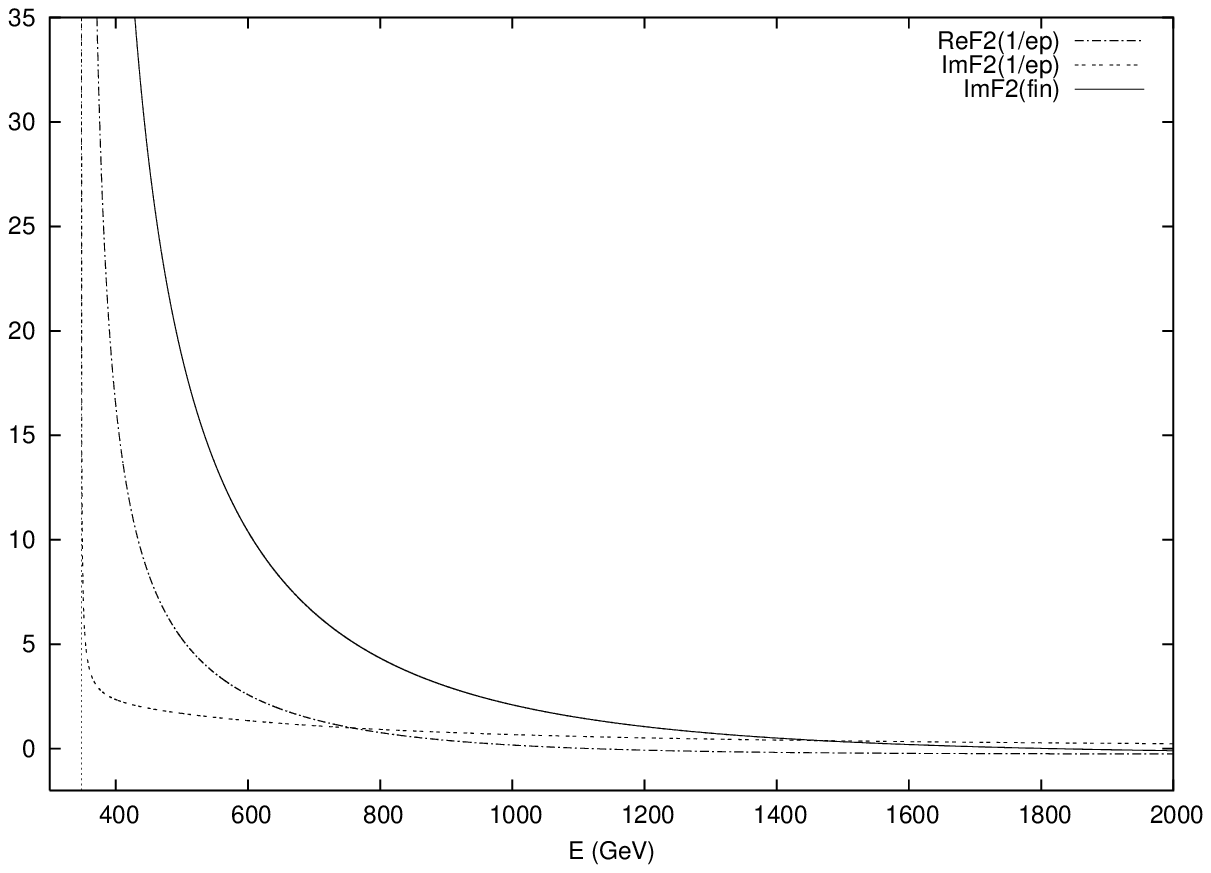}
\includegraphics[height=52mm,angle=0]{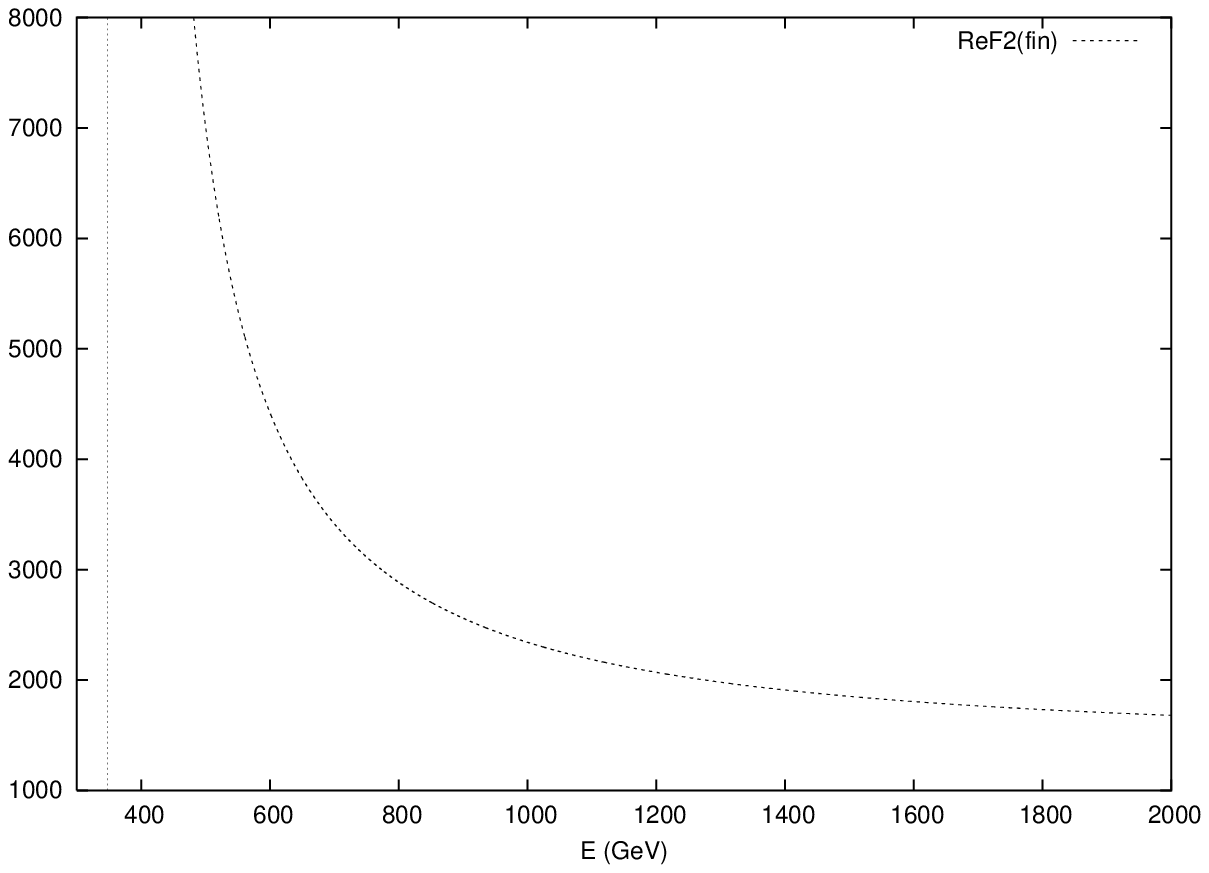}
\vspace*{-12mm}
\caption{Two-loop magnetic form factor $F_2$, in the case of top-antitop 
production.  The magnetic form factor has a simple IR pole.
}
\label{fig4}
\end{center}
\end{figure}

We present, in this short review, the numerical evaluation of the 
different parts of the form factors (real and imaginary parts of the
poles and finite parts) at one- and two-loop level above the threshold 
for the  case of the production of a pair of top-antitop quarks (one 
massive and five light flavours). The numerical evaluation is performed 
in F77 using the subroutine of \cite{Polylog3}. The color charges are set 
according to the $SU(3)$ group. The c.m. energy runs from the threshold 
value ($\sqrt{s} \sim 2m$, $m=178 \, GeV$) to 2 TeV. 
The plots are shown in Figs.~\ref{fig2}, \ref{fig3} and \ref{fig4}.

The knowledge of the exact analytical formula for the form factors allows
a perfect control on their numerical evaluation as well as on their 
analytical behaviour in all the regions of the spectrum. As an example,
consider the threshold region. The NNLO determination of the threshold 
limit of the vector form factors were already present in the literature 
\cite{hoang} as an independent calculation, performed with a different 
technique. With the method outlined above, however, this result can be 
achieved as a limiting case of the full calculation.

In fact, let us define the real and imaginary parts of the UV-renormalized 
form factors at one and two loops as:
\be
F_{i}^{(1l,2l)} = 
C(\epsilon) \, \Bigl\{
\Re \, {\mathcal F}_{i}^{(1l,2l)}
+ i \pi \, \Im \, {\mathcal F}_{i}^{(1l,2l)} \Bigr\} ,
\ee
with $i=1,2$ and $C(\epsilon)= (4 \pi)^{\epsilon} \Gamma(1+\epsilon)$. 
Moreover, consider the total cross-section for the process 
$e^{+}e^{-} \rightarrow Q \bar{Q}$ near the threshold:
\be
\sigma_{e^{+}e^{-} \to Q \bar{Q}} \! = \! \sigma^{(0)} \! \biggl[ 1 \! + 
\! \frac{\alpha_S}{2 \pi} \Delta^{(1)} \! + \! 
\left( \frac{\alpha_S}{2 \pi} \right)^2 \! \! \! \! \Delta^{(2)} 
\biggr] ,
\ee
where $\sigma^{(0)}=(2 \pi \alpha^2 e_{Q}^{2}/s) \, \beta(3-\beta^2)$ is the 
tree-level cross-section and $\beta = \sqrt{1-\frac{4m^2}{s}}$ is the quark
velocity. The one- and two-loop corrections, $\Delta^{(1)}$ and $\Delta^{(2)}$, 
can be written in terms of the form factors, up to ${\mathcal O}( \beta^2)$, 
in the following way:
\bea
\Delta^{(1)} \! \! \! \! & = & \! \! \! \! 2 \Bigl[ \Re {\mathcal F}^{(1l)}_{1}
              + \Re {\mathcal F}^{(1l)}_{2} \Bigr] \, , \\
\Delta^{(2)}  \! \! \! \! & = & \! \! \! \! 
                  \Bigl[ \Re {\mathcal F}^{(1l)}_{1}\Bigr]^2
              + 2 \, \Re {\mathcal F}^{(1l)}_{1}
	                 \Re {\mathcal F}^{(1l)}_{2} 
	      + 2 \, \Re {\mathcal F}^{(2l)}_{1} \nn\\
\! \! \! \! & & \! \! \! \! + \pi^2 \! \Bigl[ \Im {\mathcal F}^{(1l)}_{1} \Bigr]^2
	      \! + \! 2 \pi^2 \, \Im {\mathcal F}^{(1l)}_{1} 
	                 \Im {\mathcal F}^{(1l)}_{2} \nn\\
\! \! \! \! & & \! \! \! \! 	  
	      + \! \Bigl[ \Re {\mathcal F}^{(1l)}_{2} \Bigr]^2 
	      \!  \! + 2 \, \Re {\mathcal F}^{(2l)}_{2}\! 
	      + \!  \pi^2 \Bigl[ \Im {\mathcal F}^{(1l)}_{2} \Bigr]^2
	     \! .
\eea
Using the analytical expressions for the threshold limits of $\Re {\mathcal F}^{(1l)}_{1}$,
$\Im {\mathcal F}^{(1l)}_{1}$, $\Re {\mathcal F}^{(1l)}_{2}$, $\Im {\mathcal F}^{(1l)}_{2}$, 
$\Re {\mathcal F}^{(2l)}_{1}$, and $\Re {\mathcal F}^{(2l)}_{2}$, given in 
\cite{us}, the functions $\Delta^{(1)}$ and $\Delta^{(2)}$ read:
\bea
\Delta^{(1)} \!\! \! \!  & = &  \! \! \! \! C_{F} \biggl[
\frac{6 \zeta_2}{\beta} - 8 \biggr] ,
\label{delta1} 
\\
\Delta^{(2)} \! \! \! \! & = & \! \! \! \! 
C_{F}^{2} \biggl\{ \frac{12 \zeta^{2}_2}{\beta^2} 
- \frac{48 \zeta_2}{\beta} \! + \! 39 - \frac{142}{3} \zeta_2
- 4 \zeta_3\nn\\
\! \! \! \! & & \! \! \! \!  
+ \! 24 \zeta^{2}_2
+ 32 \zeta_2 \ln{2} - 16 \zeta_2 \ln{\beta} \biggr\} \nn\\
\! \! \! \! & & \! \! \! \! + C_{F} C_{A} \biggl\{ \frac{1}{\beta} 
\Bigl[ \frac{31}{3} \zeta_2 
- 22 \zeta_2 \ln{(2 \beta)} \Bigr] - \frac{151}{9}  \nn\\
\! \! \! \! & & \! \! \! \! + \frac{179}{3} \zeta_2 \! - 26 \zeta_2 \! 
 - \! 64 \zeta_2 \ln{2} \! - \! 24 \zeta_2 \ln{\beta} \biggr\} \nn\\
\! \! \! \! & & \! \! \! \! - C_{F} T_{R} N_{f} \biggl\{ \frac{1}{\beta} 
\Bigl[ \frac{20}{3} \zeta_2 \! - \! 8 \zeta_2
\ln{(2 \beta)} \Bigr] \! - \frac{44}{9}  \biggr\} \nn\\
\! \! \! \! & & \! \! \! \! 
+ C_{F} T_{R} \biggl\{ \frac{176}{9} - \frac{32}{3} \zeta_2  \biggr\} ,
\label{delta2}
\eea
where $\zeta_2$ and $\zeta_3$ are the Riemann $\zeta$ function evaluated in 
2 and 3 respectively. Eqs.~(\ref{delta1},\ref{delta2}) are in complete 
agreement with the results in \cite{hoang}.

To conclude, we presented the calculation of the two-loop QCD contributions
for the vector form factors of the photo-production of heavy quarks. By means 
of the Laporta and differential equations methods we were able to calculate
analytically the form factors, keeping the full dependence on the heavy-quark 
mass. We presented the numerical evaluation of the form factors as well as 
their threshold limit, interesting for the calculation, in this kinematical 
region, of the total cross section for the photo-production of heavy quarks. 
This result is in complete agreement with the existing result in the 
literature.

{\bf Acknowledgment}
The author wishes to thank W.~Bernreuther, T.~Gehrmann, R.~Heinesch,
T.~Leineweber, P.~Mastrolia and E.~Remiddi for their collaboration and
B. Tausk for useful discussions.

\end{document}